\documentclass[english]{llncs}

\usepackage{graphicx}
\usepackage{comment}
\usepackage{multirow}
\usepackage{amsfonts}
\usepackage{subfigure}
\usepackage{hyperref}
\usepackage{breakurl}
\usepackage{babel}
\usepackage{listings}
\usepackage{url}

\usepackage{amsthm}

\newtheorem{myexample}{Example}

\title{MITHYS: Mind The Hand You Shake} 
\author{Mauro Conti\inst{1} \and Nicola Dragoni\inst{2} \and Sebastiano Gottardo\inst{1,2}}
\institute{University of Padua, IT 
\\ \url{conti@math.unipd.it, sgottard@studenti.math.unipd.it}
\and
Technical University of Denmark, DK
\\ \url{ndra@dtu.dk, s124645@student.dtu.dk}
}

\begin{document}
\maketitle

\begin{abstract}

Recent studies have shown that a significant number of mobile applications, often handling sensitive data such as bank accounts and login credentials, suffers from SSL vulnerabilities. Most of the time, these vulnerabilities are due to improper use of the SSL protocol (in particular, in its \emph{handshake} phase), resulting in applications exposed to man-in-the-middle attacks. In this paper, we present MITHYS, a system able to: (i) detect applications vulnerable to man-in-the-middle attacks, and (ii) protect them against these attacks. We demonstrate the feasibility of our proposal by means of a prototype implementation in Android, named MITHYSApp. A thorough set of experiments assesses the validity of our solution in detecting and protecting mobile applications from man-in-the-middle attacks, without introducing significant overheads. Finally, MITHYSApp does not require any special permissions nor OS modifications, as it operates at the application level. These features make MITHYSApp immediately deployable on a large user base.

\end{abstract}

\section{Introduction}
\label{Introduction}



The spread of mobile smartphones have led web service providers to pay attention to how the users could benefit from their services, while users are on the move. 
To this end, two main approaches have been adopted. At first, providers chose to offer a mobile-shaped version of their web service, which the users could access through a mobile web browser (acting as a ``thin'' client). As an alternative, providers started to offer their services by means of native applications for each specific mobile platform (also called ``fat client'' approach). This second approach rapidly became the most popular (interested readers can refer to \cite{Charland:2011:MAD:1941487.1941504} for a thorough comparison between the two approaches). Indeed, as the number of daily activated devices grows at a relentless rate, so does the number of applications which are downloaded and available to a huge end-user base. 

An application that relies on a web service requires an active Internet connection. To gain this connection, a mobile device is typically equipped with two types of network interfaces: a 3G/4G module and a Wi-Fi module. The Wi-Fi module gives the user the opportunity of connecting a device to a wireless network created through a wireless access point. The Wi-Fi connection became more and more important, as many companies started offering free Internet access points, as an additional service for their customers. We can also find this scenario in many public infrastructures, such as libraries and universities. Unfortunately, this increasing popularity of free access points has led to new malicious attacks, based on the Man-In-The-Middle principle (from now, MITM attack). The \textit{rogue access point attack} is a typical example of how dangerous the use of a free public access point might be \cite{4455018}. As a consequence, protecting the communication in these open environments is crucial to keep user data private. This means that a mobile device must establish a secure connection with the remote server offering the needed web service. In a desktop environment, this connection lies between the web browser and the remote server. On the other hand, a mobile application is directly responsible of establishing the secure connection with the remote server, without relying on a web browser. 

Technically speaking, the most common way of establishing a secure connection is by using Secure Sockets Layer (SSL) \cite{ssl_rfc} and Transport Layer Security (TLS) \cite{tls_rfc}, two cryptographic protocols that grant endpoint authentication and network data confidentiality over a TCP connection. These protocols were also designed to prevent malicious MITM attacks against two communicating entities. The problem is that, as recently pointed out \cite{Fahl:2012:WEM:2382196.2382205}, a significant number of mobile applications often do not perform the required steps to ensure a secure communication between the communicating parties. The flowing data between the application and the server, which is supposedly private, can be intercepted by a malicious third party by performing a MITM attack. This is a known problem that affects a huge number of mobile applications, mainly due to the respective developers that underestimate the importance of a proper use of the SSL/TLS protocols. Even if the problem has been raised more than one year ago, our recent test revealed that several applications (including widely used ones, such as PayPal and Facebook) are still vulnerable. 

\begin{myexample}
\label{example_1}
Let us assume a scenario where an attacker performs a rogue access point attack, with Starbucks' free Wi-Fi service as a target. The original Starbucks' access point (AP from now on) name is ``Starbucks'', while the attacker's AP name is ``Starbucks Free''. Let us suppose Alice visits Starbucks and notices the free Wi-Fi opportunity. She sees two open access points on her Android smartphone, so she chooses a random one, the attacker's ``Starbucks Free'' in this case. Alice wants to check her PayPal account, therefore she opens the PayPal Android application, which she had used before. Since the PayPal application suffer from the above SSL usage problem, the attacker is able to intercept Alice's PayPal account data, including her personal login information. What is more, she is not aware that she is a victim of a MITM attack.
\end{myexample}

Again, given the huge number of vulnerable applications, the ``wait-and-hope'' approach is not appropriate, since it exposes the users to malicious MITM attacks until the developers release a security update. Instead, there is the need for an application-independent solution that: (i) detects the vulnerable applications; (ii) warns the user about the potential leak of sensitive data; and (iii) eventually compensates the lack of security by performing the adequate checks. Such a solution would not only secure the application-web server communication, it would also act as a security tool for mobile developers --- who want to test the security level of their applications against SSL-based MITM attacks.

\paragraph{Contribution.}
In this paper we present MITHYS (Mind The Hand You Shake), a platform independent system architecture that:
\begin{itemize}
	\item Detects mobile applications vulnerable to SSL-based MITM attacks, automatizing the detection of vulnerabilities pointed out in \cite{Fahl:2012:WEM:2382196.2382205},\cite{Georgiev:2012:MDC:2382196.2382204})
	\item Protects mobile applications (especially, vulnerable ones) from SSL-based MITM attacks, by taking care of SSL certificate validation 
	\item Gives the user full control on the vulnerable applications' behavior (e.g. the application can be blocked if vulnerable)
\end{itemize}

The MITHYS architecture is, to the best of our knowledge, the first solution that tackles the vulnerability of mobile applications to SSL-based MITM attacks \cite{Fahl:2012:WEM:2382196.2382205},\cite{Georgiev:2012:MDC:2382196.2382204}. A fully-working, end-user-ready implementation of MITHYS, namely MITHYSApp, has been developed for the Android mobile platform, which represents one of the most flexible and popular mobile OS at the time being.

Being implemented at the application level, MITHYSApp does not require mobile OS alterations nor special permissions (i.e., root access). MITHYSApp just relies on a single manual configuration performed by the user. According to the selected configuration, MITHYSApp can operate in three modes:
\begin{itemize}
	\item Automatic - detection of vulnerable applications and protection for all the installed applications, without requiring any user interaction;
	\item Selective - detection of vulnerable applications is automatic, but the user can decide whether to allow their execution or not;
	\item Manual - the user can manually select which applications must be analysed and which must be protected.
\end{itemize}

Finally, a set of experiments show the feasibility of our solution. In particular, we show that the current (non-optimized) version of MITHYSApp does not introduce a significant delay in network communication nor in the ordinary applications/OS behavior, while it effectively protects users from MITM attacks that can steal personal and sensible information.

\paragraph{Roadmap.} Section \ref{Related Work} discusses related work. Section \ref{SSL Overview} introduces the details of the security problem we solve. Section \ref{MITHYS Architecture} presents MITHYS, our solution for protecting mobile applications vulnerable to MITM attacks. Section \ref{Implementation of MITHYS} focuses on the implemetation of MITHYSApp. Section \ref{System Evaluation} evaluates our solution in terms of effectiveness and network delay. Finally, Section \ref{Conclusion} concludes the paper.

\section{Related Work}
\label{Related Work}

Today's smartphones are capable of handling different types of personal data, which most of the times can be considered sensible. As a result, smartphones security is becoming more and more a key topic in the security research community, generating a lot of studies about dangerous threats and possible solutions (as shown by the proceedings of recent top conferences on security, such as ESORICS, POLICY and CCS). Considering only the Android case and to mention only a few papers, Davi et al. \cite{davi2011privilege} presented an analysis of the privilege escalation attacks, together with some possible approaches to the problem \cite{bugiel2011xmandroid}, \cite{bugiel2012towards}. Becher et al. \cite{5958024} gave a more general security overview about the mobile smartphones environment, whereas Shabtai et al. \cite{5396322} focused more deeply in an Android security assessment. Other works focused on the direction of extending Android security features: e.g. considering Context-based access control \cite{conti2011crepe} and enforcing different modes of uses based on security profiles \cite{russello2012moses}. To mention all the papers aiming at securing Android is out of the scope of this paper. What we consider instead important to point out is that, although this increasing research effort, a significant work has still to be done in order to secure smartphone platforms. This is proved by the huge vulnerability recently discovered regarding the use of the SSL cryptographic protocol.

Various misuses of the SSL protocol are spread both in the desktop environment and in the mobile environment, exposing private data (potentially sensible) to malicious attacks. In particular, Georgiev et al. \cite{Georgiev:2012:MDC:2382196.2382204} analysed the SSL usage across various environments, only to find out that this protocol's implementation is ``completely broken in many security-critical applications and libraries''. Meanwhile, Fahl et al. \cite{Fahl:2012:WEM:2382196.2382205} analysed the SSL usage on 13,500 Android applications, and found out that a large percentage of them suffer from SSL vulnerabilities, which expose them to dangerous man-in-the-middle attacks. To add it up, some of these applications (such as PayPal and Facebook) are very popular, covering up to 185 million users. Both studies just gave some advices to developers, but did not mention any solution to the SSL usage problem.

SSL misuse vulnerabilities have been also considered in the literature. For example, the work in \cite{Benton:2011:SPD:2179298.2179365} shows an approach to detect SSL-based man-in-the-middle-attacks. However, this approach is designed for desktop web browsers, so it is not suitable for the setting of mobile applications that we are considering in this work. Furthermore, a simple MITM attack towards the third-party server proposed in \cite{Benton:2011:SPD:2179298.2179365} completely invalidates their protection mechanism. This problem is also acknowledged by the authors in their work.

Despite the size of the problem, the SSL usage vulnerability problem for mobile applications is still out there, threatening millions of users and their private data. We will focus on this problem in the next Section. 








\section{The Problem: Validating SSL Certificates}
\label{SSL Overview}

Nowadays Internet browsers, electronic mail clients, instant messaging clients, and nearly every entity that needs a secure communication to a remote service are using SSL and TLS, two standard cryptographic protocols that perform network data encryption and endpoint authentication over a TCP connection. An SSL secure communication begins with an operation called \textit{handshake}, in which the server is authenticated by the client (and viceversa, eventually). After that, these two entities agree on a common cryptographic material, used to begin the encrypted communication. This flow can be roughly summarised as follows (we are not considering the client authentication steps, which are optional):

\begin{enumerate}
	\item The client contacts the server, and they exchange some preliminary parameters, among which the certificates (the client's certificate is optional, therefore often missing); the exchanged parameters are called context of a SSL session.
	\item The client authenticates the server by using the information obtained in the previous step, especially the server's certificate; for a secure session to be established, the server must be successfully authenticated by the client (either implicitly or explicitely).
	\item The client, thanks to the previous information exchange, creates a pre-master secret, encrypted with the server's public key obtained from the server's certificate, and sends it to the server.
	\item The server decrypts the message and uses the pre-master secret to compute the master secret while the client does the same. 
	\item Using the master secret, both the client and the server generate the so called session keys, that will be used to communicate securely.
	\item The communication starts as the client sends the first encrypted message.
\end{enumerate}

There is a slight problem on the second point of the above flow. The client must authenticate the server in order to be sure that it is communicating with the right server and not with, for instance, a malicious one which is faking its identity (a typical MITM situation). This is mostly done by thoroughly checking the server's SSL certificate fields (e.g., expiration date, issuer, signature). 

\begin{myexample}
\label{example_2}

Continuing the scenario described in Example~\ref{example_1}, let us suppose Alice is using PayPal's Android application (PayPalApp), which needs to communicate with PayPal's remote server (PayPalServer). However, the attacker (MITM) is able to intercept the ingoing and outgoing traffic of PayPalApp. The following steps are performed as part of the SSL handshaking process: 
\begin{enumerate}
	\item PayPalApp queries PayPalServer for its X.509 certificate (which contains PayPalServers's public key).
	\item MITM intercepts PayPalApp's request and asks PayPalServer for its certificate pretending she is PayPalApp; PayPalServer sends its certificate to MITM.
	\item MITM now generates a fake X.509 certificate containing MITM's public key instead of the PayPalServer's one; MITM also makes this fake certificate look like PayPalServer's one, then sending it back to PayPalApp.
	\item Depending on how strict are PayPalApp's checks against MITM's certificate, PayPalApp will eventually think that she's talking to PayPalServer.
	\item At this point, MITM can intercept the plain text of every message (i.e., MITM can easily decrypt the messages) PayPalApp sends to PayPalServer and viceversa, but she is undetected.
\end{enumerate}
\end{myexample}

In Example~\ref{example_2}, \emph{PayPalApp} performs very poor checks against MITM's certificate (e.g., it might not check the issuer name of the certificate, therefore not recognizing a MITM attack). As a result, Alice is not able to detect that the communication with \emph{PayPalServer} is not secure at all, allowing MITM to intercept all the available data. It is important to stress that this is not just a toy example, we have actually developed a demo implementing this specific attack.

It is clear by now that the key point of this procedure consists in validating the server's certificate in a proper way. Since many mobile applications do not perform this step correctly, exposing the end-user to dangerous MITM attacks, our solution focuses on solving this specific problem.

\section{MITHYS: Mind The Hand You Shake}
\label{MITHYS Architecture}

In this section, we present MITHYS (Mind The Hand You Shake), a system designed to detect potentially MITM-vulnerable applications, and to compensate the lack of security by protecting applications from MITM attacks. To the best of our knowledge, MITHYS represents the first solution that tackles the MITM vulnerability of mobile applications by taking on the security checks required to establish a proper secure connection. For space limitation, we omit details on MITHYS user interface and configuration. Instead,  we focus on the core of MITHYS and we describe it from a system point of view, focusing on its architecture, its implementation (Section~\ref{Implementation of MITHYS}) and its evaluation (Section~\ref{System Evaluation}).

The main idea behind MITHYS is to act as a friendly MITM on the mobile device. Every time a ``new'' application (an application which has not been tested yet) requests a resource via the HTTP over the SSL protocol (from now on, HTTPS requests), the MITHYS system tries to act as a man-in-the-middle, forging a fake ad-hoc SSL certificate for the application. If the application is not vulnerable, it will immediately block the communication; otherwise (the application is vulnerable), the communication will proceed normally, as if there is no third party between the application and the remote server. In both scenarios, MITHYS is able to protect the application from potentially malicious MITM attacks by performing additional checks on the SSL connection (Section~\ref{MITHYS Workflow}).

\begin{figure}[t]
	\centering
	\fbox{
	\includegraphics[scale=0.6]{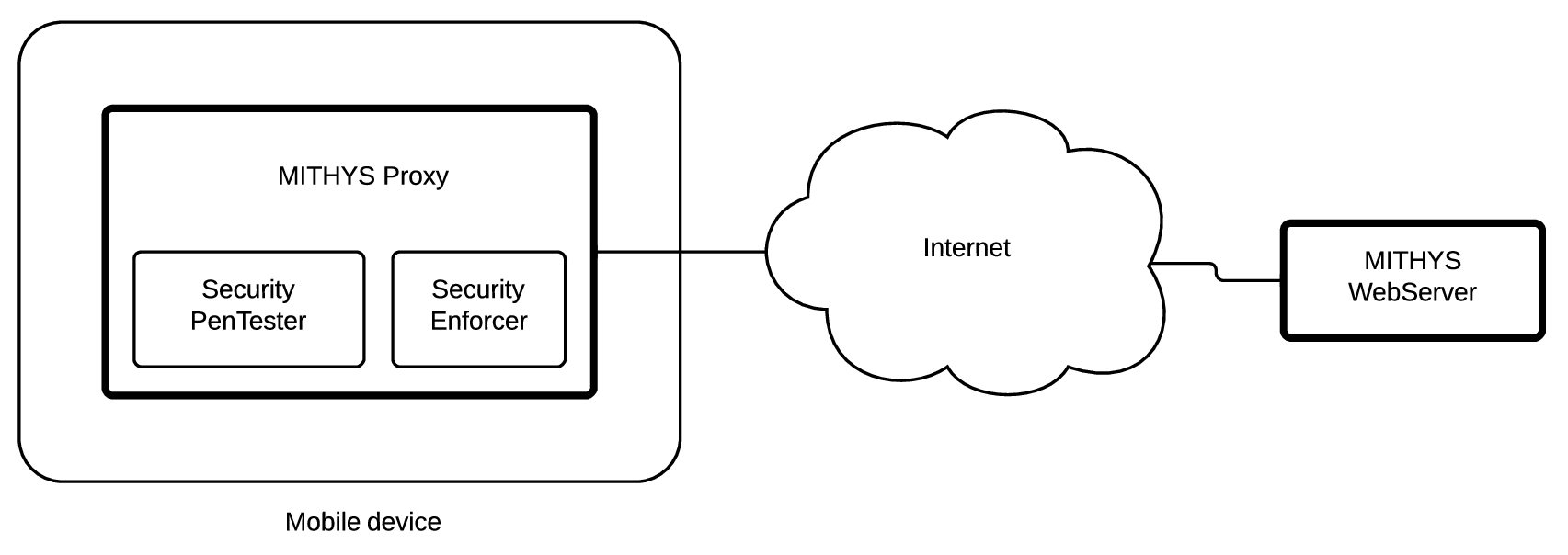}}
	\caption{The MITHYS high-level architecture}
	\label{fig:high_level_architecture}
\end{figure}

An high-level overview of the MITHYS architecture is shown in Figure~\ref{fig:high_level_architecture}. At a macroscopic level, there are two main components, highlighted in the figure by thicker borders. The first one is called \textit{MITHYS Proxy}, a proxy-based mobile application that runs on the mobile device. The second one is called \textit{MITHYS WebServer}, a remote web server hosted and reachable through the Internet. 

We now describe the two key components of MITHYS: MITHYS WebServer (Section~\ref{MITHYS WebServer}) and MITHYS Proxy (Section~\ref{MITHYS Proxy}). Then, in Section~\ref{MITHYS Workflow} we describe how the overall system works.



\subsection{MITHYS WebServer}
\label{MITHYS WebServer}
This component acts as a trusted party for the solution. It features only one servlet, whose purpose is to retrieve the SSL certificates chain (typically in the X.509 standard) of the URL passed as an argument; then, it serializes the chain in a proper way and returns it as a result. This servlet is only reachable via HTTPS, meaning that it has a SSL certificate associated to it. This is a key point of the whole architecture. This SSL certificate is self-signed, i.e. generated from the root certificate of our private Certification Authority (i.e., \textit{MITHYS CA}). Since we have access to the original certificate, we can use its information to add an extra layer of security against MITM attacks, as we discuss in Section~\ref{MITHYS Proxy}. Finally, we underline that we do not consider this component as a possible target for attacks, mainly because (i) it can be hosted on highly secure cloud services (e.g., Google Compute Engine) and (ii) it is easier to protect this single component rather than protecting millions of user devices with an highly variable set of installed applications. However, in order to prevent Denial-of-Service (DoS) attacks, we recommend the redundancy approach, by means of a MITHYS WebServer pool.

\subsection{MITHYS Proxy}
\label{MITHYS Proxy}


This represents the main component of the architecture. Its main purpose is to receive all the HTTPS requests coming from the applications installed on the mobile device, and to pass the information back and forth between the application and its associated web server. It can also strengthen the applications' security by performing additional checks (as detailed later in this section) on the SSL connection. In order to fulfill its tasks, it features two independent modules (see Figure~\ref{fig:high_level_architecture}): \textit{Security PenTester} and \textit{Security Enforcer}.

\paragraph{Security PenTester.}
\label{mithys_sp}
This module is the component which represents the actual MITM. It impersonates the original remote server by forging a fake SSL certificate for the mobile application. It also contacts the original remote server, pretending to be the application itself. If Security PenTester is able to establish a secure connection with the application (that is to say, the application accepts the fake SSL certificate), it acknowledges that the application is vulnerable. Otherwise, we can only have some degree of confidence that the application is not vulnerable, while it could be actually vulnerable in other circumstances. This module runs continuously, so every application is basically tested every time it issues an HTTPS request. Since we want \textit{``PenProof''} applications (i.e., applications that are not vulnerable to the PenTester) to be excluded from further security tests, an effective approach consists in adding them to a whitelist: every application on that list avoids the Security PenTester module, but may still be strengthened by the Security Enforcer module.

We want to point out that the use of a whitelist is actually mandatory. A PenProof application that receives a fake SSL certificate for an HTTPS request will terminate the connection immediately, therefore not working correctly. As a consequence, the MITHYS system needs to be aware of the already (successfully) tested applications, so that we do not hinder their normal operations.

\begin{figure}[tb]
	\centering
	\fbox{
	\includegraphics[scale=0.6]{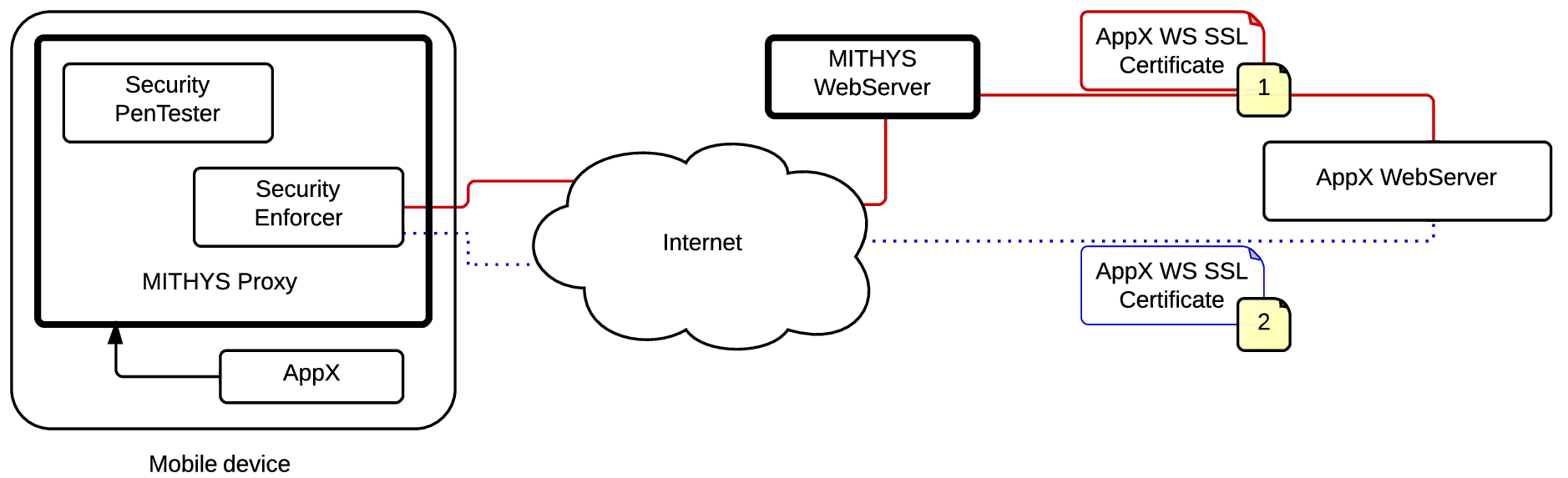}}
	\caption{The MITHYS Security Enforcer interaction scheme.}
	\label{mithys_se_interaction_scheme}
\end{figure}

\paragraph{Security Enforcer.}
\label{mithys_se}
This module performs additional checks on the SSL connection to the remote server in place of the mobile application. More specifically, given the HTTPS request issued by AppX (an installed application), this module performs the following operations (illustrated in Figure~\ref{mithys_se_interaction_scheme}):

\begin{itemize}
	\item Issues an HTTPS request to the MITHYS WebServer, in order to retrieve the SSL certificates chain associated to the URL of the application's HTTPS request (Step 1 in the figure);
	\item Retrieves the SSL certificates chain associated to the URL of the HTTPS request (Step 2 in the figure);
	\item Compares the two certificates chains. Each certificate of one chain is compared to the respective certificate of the other chain. This is done by checking if the signatures of the two certificates correspond.
\end{itemize}

If the certificates contained in the two chains do not match, it means that a MITM attack might be in place. On the other hand, if the two chains have a 1:1 match, we can be sure that no SSL-based MITM attack is being held at that time. This assumption is based on the fact that the HTTPS request to the MITHYS WebServer is MITM-proof. To achieve such requirement, since the SSL certificate of our MITHYS WebServer is known a priori, we can store it on a keystore and embed it in our MITHYS Proxy mobile application. So, when the HTTPS request to the MITHYS WebServer is issued, the obtained SSL certificate is matched against our keystore: any failure will invalidate the certificates chains comparison, indicating an ongoing MITM attack of some kind.
It is worth pointing out that an application which has passed the Security PenTester's controls might still be monitored by the Security Enforcer (e.g., as an extra security measure for the user). What is more, Security Enforcer only sends to MITHYS WebServer the URL of the original HTTPS request, without transmitting any sensitive information of the user.

\subsection{MITHYS Workflow}
\label{MITHYS Workflow}

\begin{figure}[h!]
	\centering
	\fbox{
	\includegraphics[scale=0.60]{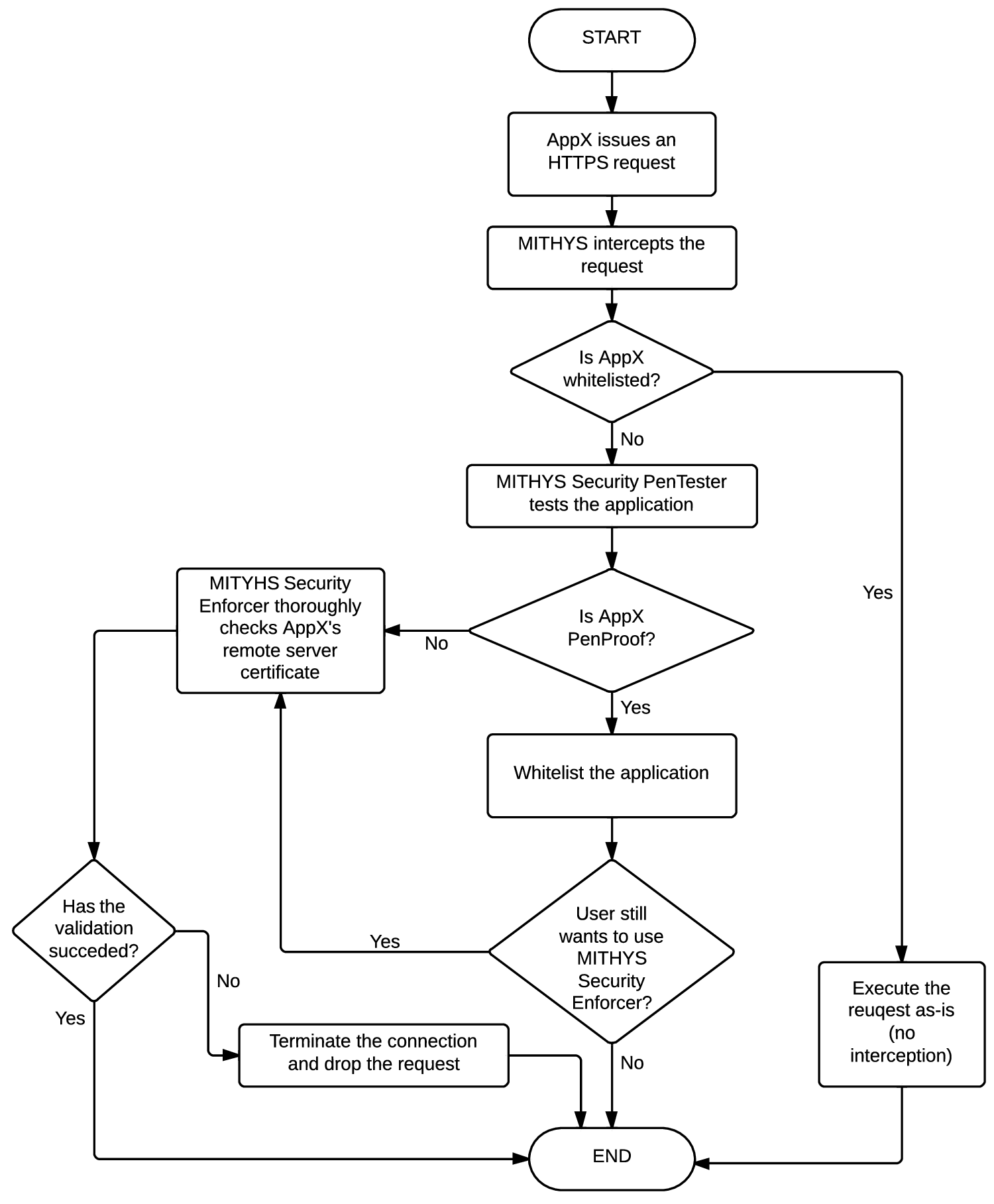}}
	\caption{Workflow of the MITHYS architecture with the AppX mobile application.}
	\label{fig:mithys_architecture_appx_workflow}
\end{figure}

In order to better understand how the overall MITHYS system works, Figure~\ref{fig:mithys_architecture_appx_workflow} shows a simplified workflow of a generic scenario where the mobile application AppX issues an HTTPS request (e.g., to \textit{https://www.appx.com/api/login}). The request is intercepted by our MITHYS Proxy, that checks whether the application has ever been whitelisted. If not, Security PenTester tries to act as a MITM and determines if AppX is aware of a third entity between AppX's remote server and itself. If the application is aware of the MITM, it is whitelisted: each subsequent HTTPS request coming from that application will be executed as is, without any interception. Otherwise, Security Enforcer is activated in order to prevent any malicious MITM attacks. Again, note that even a whitelisted application might take advantage of the latter module, if specified by the user.

\begin{myexample}
\label{example_3}
Back to our running example, let us consider Example~\ref{example_2} to show the workflow of MITHYS with PayPal's Android application. The key assumption is that Alice is using a MITHYS implementation on her smartphone. Alice starts the PayPalApp, which in turn issues HTTPS requests to the PayPalServer. These requests are intercepted by MITHYS' Security PenTester (PenTester from now on). PenTester retrieves the list of whitelisted applications to check if PayPalApp is among those. The whitelist is initially empty, so PenTester acts as a SSL MITM and forges a fake SSL certificate. PayPalApp, as we show in Section~\ref{Vulnerability Detection}, is vulnerable to this attack, so it accepts the certificate. Now that PenTester has acknowledged that PayPalApp is vulnerable, it reports this information to the MITHYS' Security Enforcer module (Enforcer from now on). Enforcer must now protect PayPalApp from actual MITM attacks by performing the steps described in Section~\ref{mithys_se}. What is more, Enforcer will protect all the future PayPalApp's HTTPS requests. 
\end{myexample}

\begin{myexample}
\label{example_4}
We reconsider Example~\ref{example_3}, but we assume that this time Alice wants to use the Twitter application, which is not vulnerable to SSL MITM attacks (Section~\ref{Vulnerability Detection}). Again, Alice is using a MITHYS implementation. Alice starts TwitterApp, PenTester intercepts the HTTPS requests to TwitterServer and tries to act as a SSL MITM for TwitterApp. The latter is not vulnerable, so it will reject the fake SSL certificate and abort the current operation. Now PenTester knows that the application is secure, so it adds TwitterApp as a new whitelist entry. TwitterApp can operate without the Enforce protection, but the user might want to be protected anyway. If this is the case, Enforcer will protect all the future TwitterApp's HTTPS requests. Otherwise, it will simply forward the HTTPS requests/responses between TwitterApp and TwitterServer.
\end{myexample}

\section{Implementation of MITHYS: MITHYSApp}
\label{Implementation of MITHYS}

This section discusses our implementation of MITHYS, namely the MITHYSApp Android application which acts as the MITHYS Proxy component. The MITHYS WebServer consists in a Micro Instance of Amazon's Elastic Compute Cloud Web Services (AWS EC2) \cite{amazon_ec2}: a continuously running Apache Tomcat instance serves an HTTPS-only Java servlet called \verb+GetSSLCertificate+.

\subsection{The MITHYSApp WebServer}
\label{The MITHYSApp WebServer}

MITHYSApp WebServer implements the MITHYS WebServer component. It is hosted on Amazon Elastic Compute Cloud (Amazon EC2) \cite{amazon_ec2} as part of the Amazon Web Services. A Micro Instance of the EC2 cloud, which we can consider as a proper Virtual Private Server (VPS), runs the Apache Tomcat web server and servlet container. There is only one servlet, called \verb+GetSSLCertificateServlet+ that takes in input two arguments: the first one is the target URL, the second one is the HTTP method that should be used to invoke that URL. This servlet simply issues an HTTPS request to the target URL (accordingly to the HTTP method) and retrieves the SSL certificates chain associated to that URL. The Base64 serialization of the chain is returned as a JSON-formatted result. Please note that this servlet is only available via HTTPS, and it uses an SSL certificate generated from our MITHYS Certification Authority (MITHYS CA) in order to prevent MITM attacks against our MITHYSApp application.

\subsection{The MITHYSApp Android Application}
\label{The MITHYSApp Android Application}


MITHYSApp is an Android app that implements the MITHYS Proxy component. It relies on the open source Android library SandroProxyLib\footnote{\url{https://github.com/SandroB/sandrop/tree/master/projects/SandroProxyLib}}, which is based in turn on the OWASP WebScarab project, that offers a working-out-of-the-box proxy for Android. What is more, it behaves as the MITHYS Security PenTester by default due to the fact that, every time it receives a new HTTPS request, it acts as a MITM and forges ad-hoc fake certificates. These certificates are generated from the MITHYS CA, and their hostname matches the hostname of the target server, looking similar to the original ones. From now on we will use also the term ``proxy'' to refer to the proxy part of this library. While not requiring any special permission or OS modifications, MITHYSApp requires the installation of the MITHYS CA certificate and the setup of the proxy address for the current Wi-Fi connection. MITHYS guides the user in both these steps, both performed only once at installation time.

\paragraph{Security PenTester.} We had to modify and to extend the SandroProxyLib library in order to implement the above component correctly. First of all, given an intercepted HTTPS request, we need to know which application generated it: in terms of Java objects, we only have a \verb+Socket+ instance that represents the connection between the application and the proxy, of which we only know the port. But, since Android is a Linux-based OS, we can read the content of the \textit{/proc/net/tcp} (or \textit{/proc/net/tcp6} if an IPv6 address is available) file that maps all the active sockets to their Unix processes: in this way we know which port is being used, so we can obtain the UID of the process which is using that port. This information, together with the \verb+PackageManager.getPackagesForUid(uid)+ method provided by Android, offers us the possibility of knowing which application issued the HTTPS request given just the port of its \verb+Socket+ object. To the best of our knowledge, this is the only technique available at the time being, so we created a small and useful Android library\footnote{\url{https://github.com/dextorer/AndroidTCPSourceApp}} which eases this process for the developer.
Another modification to the proxy library consisted in introducing the whitelisting mechanism, so that each time an installed application refuses to establish a secure connection with the proxy (that is, the SSL handshake phase between our proxy and the application cannot be completed) it communicates the non-vulnerable application to MITHYSApp. To do so, an \verb+AppDescriptor+ object containing package name, application version and requested URL is created and sent to the running instance of MITHYSApp. The latter receives the \verb+AppDescriptor+ object and inserts its values on a local SQLite database. This database must be encrypted in order to prevent manual tampering, so we used a custom Android library called SQLCipher\footnote{\url{https://guardianproject.info/code/sqlcipher/}} to provide ``transparent 256-bit AES encryption of database files''. In addition, for each new HTTPS request the proxy checks if the application who issued it has been whitelisted before, by querying the SQLite database: if so, no interception is made and the proxy simply passes the data back and forth between the whitelisted application and the remote server.
In addition, in order to prevent alterations to the local MITHYS keystore, we invoke a JNI-compiled library that checks the current Java package name and the keystore size. Thanks to this approach, any attempt to (i) replace the native library, to (ii) modify the Java code of MITHYSApp or even to (iii) replace the keystore will lead to a non working application.

\paragraph{Security Enforcer.} In order to implement the Security Enforcer module, we had to extend the SandroProxyLib library so that, every time a vulnerable application issues an HTTPS request, the proxy performs the following steps:

\begin{enumerate}
	\item Retrieves the SSL certificates chain associated to the URL of the HTTPS request.
	\item Issues an HTTPS request to the MITHYSApp WebServer, in order to retrieve the SSL certificates chain associated to the URL of the application's HTTPS request.
	\item Compares the two certificates chains, as described in Section~\ref{mithys_se}.
\end{enumerate}

If no MITM attack is in place, the comparison will succeed and the HTTPS request will be issued without further ado. If a MITM attack is in place, the HTTPS request issued towards the MITHYSApp WebServer will simply fail (as we explained in Section~\ref{mithys_se}). A smarter attacker might decide not to intercept the HTTPS requests addressed to our MITHYSApp WebServer: but this won't prevent our Security Enforcer module from detecting a MITM attack, since the two certificates chains are still compared one against the other.

\section{System Evaluation}
\label{System Evaluation}

In this section, we present a set of tests that assess the performance impact of the MITHYS approach and determine its ability to successfully detect vulnerable applications. More specifically, we want to show that, although MITHYS requires additional HTTPS requests in order to protect the mobile device from MITM attacks, the user is not dramatically affected by this overhead. First, we will analyse the effectiveness of MITHYSApp's vulnerability detection in Section~\ref{Vulnerability Detection}. Then, in order to determine the additional overhead, we will discuss our test method in Section~\ref{Experimental Setting} and the results in Section~\ref{Network Overhead}.

\subsection{Vulnerability Detection}
\label{Vulnerability Detection}

In their analysis, Fahl et al. \cite{Fahl:2012:WEM:2382196.2382205} manually audited some of the most popular Android applications, in order to test their vulnerability to SSL-based MITM attacks. We manually tested the same set of applications (that, in the meantime, could have been updated, fixing this MITM vulnerability) against MITHYSApp, therefore evaluating the capability and the accuracy of detecting vulnerable applications. We show our results in Table~\ref{table:mithysapp_detection_results}. The results show that MITHYSApp is able to successfully detect vulnerable applications (according to Fahl et al.'s findings). MITHYSApp is also consistent with the results in \cite{Fahl:2012:WEM:2382196.2382205} in detecting \textit{Twitter} and \textit{Voxie Walkie Talkie} as non vulnerable.

\vspace{-0.2cm}
\begin{table}[!h]
\centering
	\subfigure{
		\begin{tabular}{l*{6}{c}r}
			Application     	& Test result \\
			\hline
			Amazon MP3			& $\times$ \\
			Chrome				& $\times$ \\
			Dolphin Browser HD	& $\times$ \\
			Dropbox	 			& $\times$ \\
			Ebay	        	& $\times$ \\
			Expedia Bookings    & $\times$ \\
			Facebook Messenger 	& $\times$ \\
			Facebook			& $\times$ \\
			Foursquare			& $\times$ \\
			GMail				& $\times$ \\
		\end{tabular}
	}
	\subfigure{
		\begin{tabular}{l*{6}{c}r}
			Application     	& Test result \\
			\hline
			Google Play Store	& $\times$ \\
			Google+				& $\times$ \\
			Hotmail				& $\times$ \\
			Instagram			& $\times$ \\
			OfficeSuite Pro 6	& $\times$ \\
			PayPal				& $\times$ \\
			Twitter				& \checkmark \\
			Voxie Walkie Talkie	& \checkmark \\
			Yahoo! Messenger	& $\times$ \\
			Yahoo! Mail			& $\times$ \\
		\end{tabular}
	}
	\vspace{3mm}
	\caption{MITHYSApp results in detecting apps safe from SSL-based MITM attacks. (\checkmark) indicates that the app is safe; ($\times$) means that the app is vulnerable.}
	\label{table:mithysapp_detection_results}
\end{table}

\vspace{-1.3cm}
\subsection{Experimental Setting}
\label{Experimental Setting}

We have tested MITHYSApp with three of the most popular Android applications. These application belong to different categories of Google's Play Store, and represent three different important aspects that a typical mobile user is interested to: social networking, finance checking, cloud storage access. In particular, the applications we considered are: Facebook\footnote{\url{https://play.google.com/store/apps/details?id=com.facebook.katana}} (social networking service), PayPal\footnote{\url{https://play.google.com/store/apps/details?id=com.paypal.android.p2pmobile}} (global e-commerce business allowing online payments and money transfers), and Dropbox\footnote{\url{https://play.google.com/store/apps/details?id=com.dropbox.android}} (web-based file hosting service).

In our tests we considered two operations common to all the applications listed above: \textit{login} and \textit{logout}. These operations are very network-intensive, hence representing a perfect test scenario for MITHYSApp.
As main tool for testing, we used \verb+monkeyrunner+ \cite{monkeyrunner_doc}. This tool allows interacting (e.g., pressing buttons, typing text) with an Android device by writing a simple Python script and running it via Android Debug Bridge (\verb+adb+\footnote{\url{http://developer.android.com/tools/help/adb.html}}). We wrote three scripts, one for each considered application. Each script basically performs these operations:

\begin{enumerate}
	\item Connects to the Android device;
	\item Opens the Android logcat in a subprocess (more on this later);
	\item Starts the application's login activity;
	\item Enters the credentials for a valid account;
	\item Presses the login button and saves the current time on a variable called \verb+LoginStartTime+;
	\item Monitors the logcat in order to see when the main activity of the application is displayed - as soon as this happens, it saves the current time on the \verb+LoginEndTime+ variable;
	\item Calculates the login time as (\verb+LoginEndTime+ - \verb+LoginStartTime+);
	\item Executes a number of actions in order to start the logout procedure; as soon as the logout button is pressed, it saves the current time on \verb+LogoutStartTime+;
	\item Monitors the logcat in order to see when the login activity of the application is displayed - as soon as this happens, it saves the current time on \verb+LogoutEndTime+;
	\item Calculates the logout time as (\verb+LogoutEndTime+ - \verb+LogoutStartTime+);
	\item Prints the two results.
\end{enumerate}

We want to focus for a moment on the use of the \verb+logcat+ \cite{logcat_doc}. This tool allows the developer to collect and view the log messages, both coming from the Android OS and from the installed applications. We used specific \verb+logcat+ messages to determine the end of each operation (login and logout). Every time that the system displays a particular activity of the application (i.e., the main activity after the login, the login activity after the logout), we are sure that the considered operation has ended. This approach leads to reliable and repeatable tests, whereas it does not pollute the tests results at all.

\subsection{Network Overhead}
\label{Network Overhead}

\begin{figure}[t]
\centering
	\fbox{
	\subfigure[Login]
	{\includegraphics[width=5.95cm]{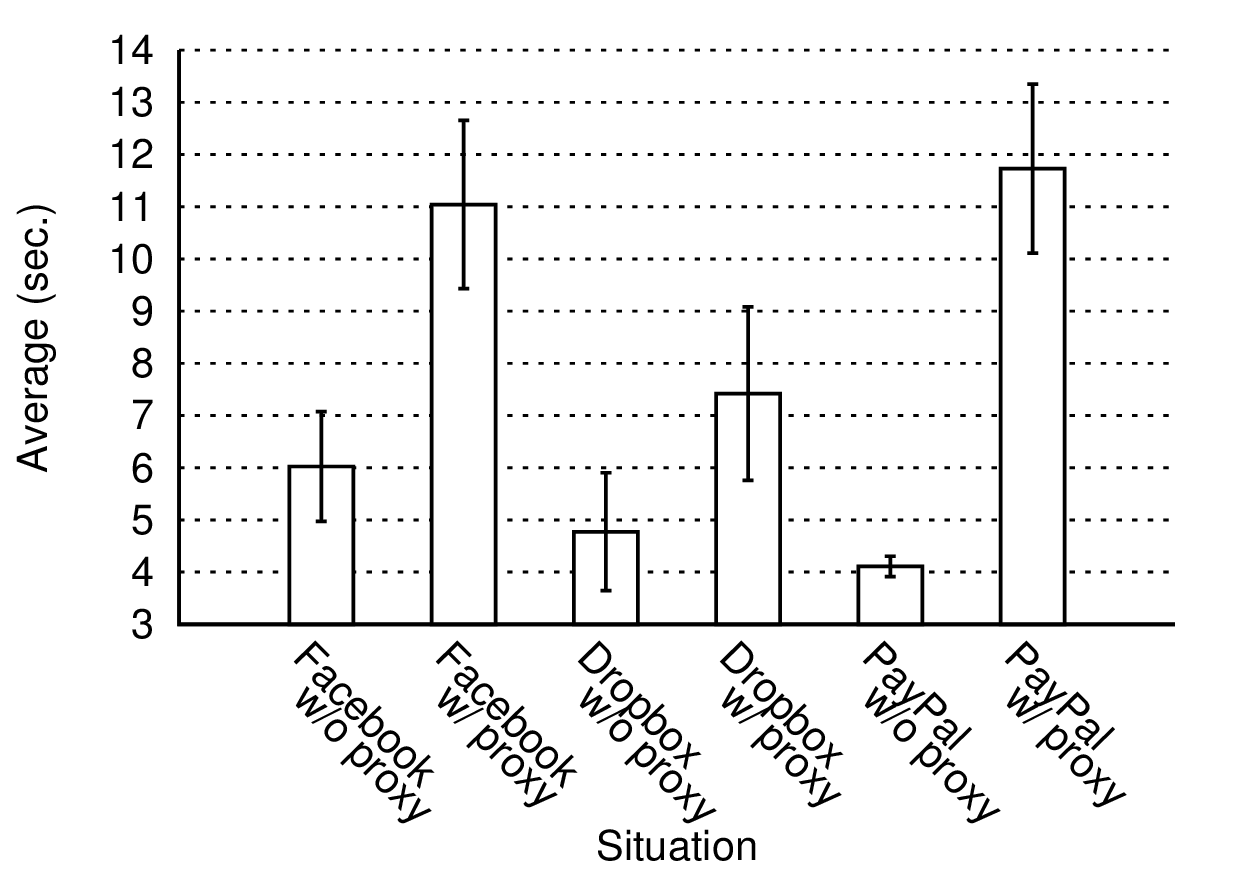}\label{fig:overhead-login}}
	\hspace{-3mm}
	\subfigure[Logout]
	{\includegraphics[width=5.95cm]{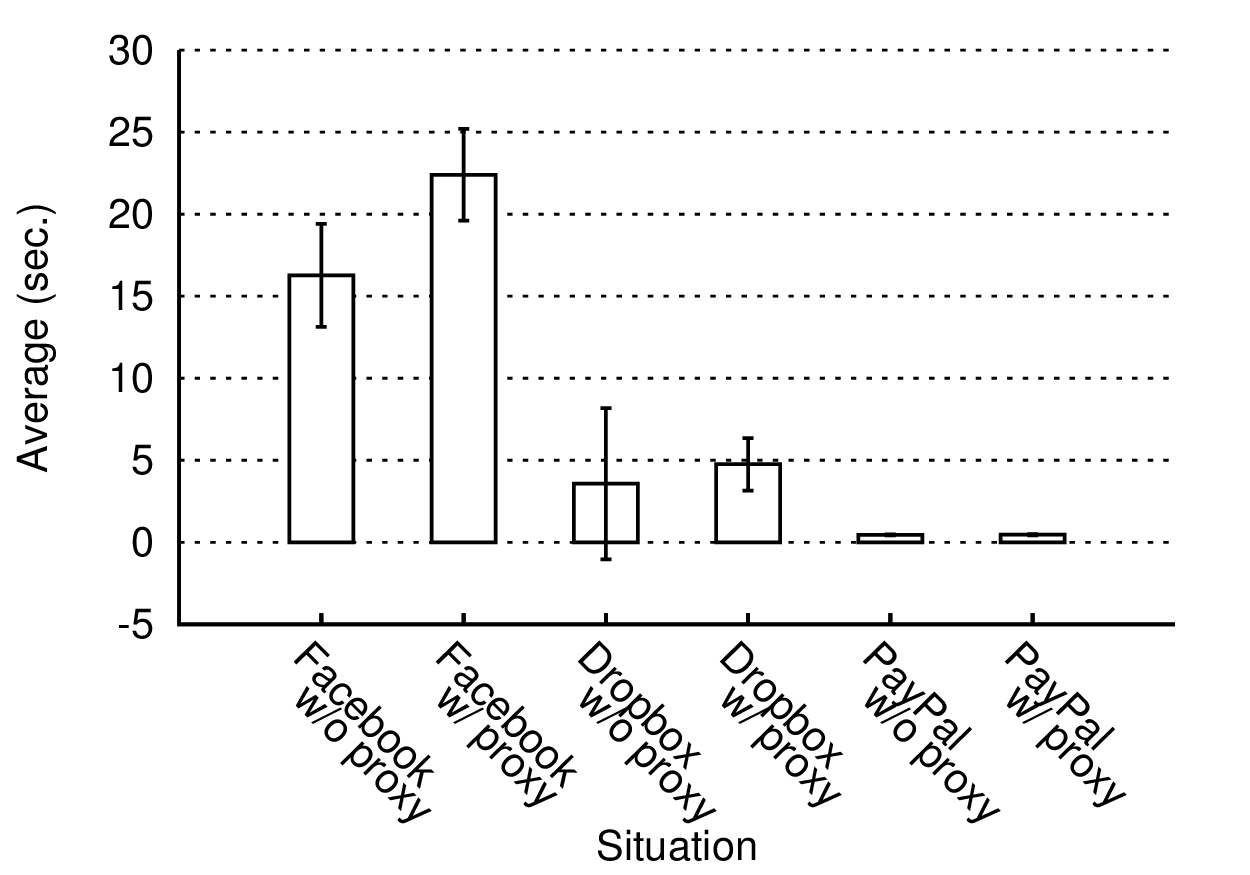}\label{fig:overhead-logout}}}
	\caption{MITHYS: time overhead for representative applications.}
	\label{fig:time-overhead}
\end{figure}

The results of our experiments are reported in Figure \ref{fig:time-overhead}. In particular, Figure~\ref{fig:overhead-login} and Figure \ref{fig:overhead-logout} represent the overhead for the login and logout operation, respectively. We can observe that the average delay added by using MITHYSApp is approximately five seconds. Since this value is almost constant for each of the considered situations, the delay is more likely to be noticed by the user for shorter operations. The two figures show a higher delay in using MITHYSApp for both the login and the logout operations. This overhead is not suprising though, becuase MITHYSApp needs to issue additional network requests in order to protect mobile applications from MITM attacks. If we consider Facebook, the introduced delay for the login operation is about 55\%, whereas for the logout operation it is about 33\%.

There is an important point here we want to stress. While the current version of MITHYSApp is a fully-working implementation, we need to consider that it has not yet been optimised, both in terms of certificate caching and in terms of network performances. As a consequence, the values that emerged from the tests can be considered as an upper bound for the additional delay, which in some situations may be indeed noticeable by the user. We believe that, by properly optimising our implementation, we can reduce the five seconds average delay to a value of three or even two seconds.
Another aspect that we have to take into account is that MITHYSApp is able to prevent MITM attacks that usually are performed nearby free Internet access points. Therefore, the user should take advantage of it while she is connected to a wireless access point, whereas it could be deactivated in other less attack-prone circumstances.

\section{Conclusion}
\label{Conclusion}

In this paper we have addressed a SSL vulnerability that has been recently shown affecting a base of many millions of users of mobile devices. To solve this problem, we have proposed MITHYS, a system for mobile devices which is able to protect mobile applications from SSL vulnerabilities. The architecture of MITHYS is light and feasible for several mobile platforms. To support this claim, we implemented MITHYSApp, i.e., MITHYS for Android. In particular, we implemented MITHYSApp at the application level, thus facilitating the spread of our solution and its installation on Android-powered mobile devices. We decided to focus on the Android platform mostly due to its popularity and flexibility. However, we have reasons to believe that mobile applications for Apple devices (e.g., iPhone, iPad) are just as vulnerable as the ones available for Android. For example, Thampi \cite{mitmproxyandpath} was able to perform an SSL-based MITM attack to analyse the Path iOS application, discovering an illegitimate upload of the user's contacts to Path's servers. As a consequence, Path released a security update to its application, acknowledging the problem \cite{thepathfact}.

The results of our experiments showed that MITHYSApp has a limited overhead that even if noticeable, we believe being accepted by users when effectively protecting them from man-in-the-middle attacks aiming at stealing personal and sensible information. MITHYSApp represents a first (though fully working) implementation of the MITHYS system. Therefore, its performances can be vastly improved by adding advanced caching mechanisms. While the delay introduced by using MITHYSApp is still acceptable, we estimate that it can be further reduced by at least two seconds.


\bibliographystyle{abbrv}

\bibliography{reference_cut}


\end{document}